# AB Levitrons and their Applications to Earth's Motionless Satellites

**Alexander Bolonkin**
C&R, 1310 Avenue R, #F-6, Brooklyn, NY 11229, USA.
(718) 339-4563, aBolonkin@juno.com, aBolonkin@gmail.com, http://Bolonkin.narod.ru

## Abstract

Author offers the new and distinctly revolutionary method of levitation in artificial magnetic field. It is shown that a very big space station and small satellites may be suspended over the Earth's surface and used as motionless radio-TV translators, telecommunication boosters, absolute geographic position locators, personal and mass entertainment and as planet-observation platforms. Presented here is the theory of big AB artificial magnetic field and levitation in it is generally developed. Computation of three macro-projects: space station at altitude 100 km, TV-communication antenna at height 500 m, and multi-path magnetic highway.

## Introduction

**Brief history**. The initial theory of levitation-flight was developed by the author during 1965 [1]. Theory of electrostatic levitation and artificial gravity for spaceships and asteroids was presented as paper AIAA-2005-4465 in 41st Propulsion Conference, 10-13 July 2005, held in Tucson, AZ, USA [2]. The related idea and theory extends from the author's work "Kinetic Anti-Gravitator" [3] presented as paper AIAA-2005-4504 in 41st Propulsion Conference. The work "AB Levitator and Electricity Storage" [4] was presented as paper AIAA-2007-4612 to 38th AIAA Plasma dynamics and Lasers Conference in conjunction with the16th International Conference on MHD Energy Conversion on 25-27 June 2007, Miami, USA. (See also http://arxiv.org search "Bolonkin").

The given work underwent further development and application of the above-cited works. That allows an estimate of the parameters of low-altitude stationary satellites, space stations, communication marts and cheap multi-path highway for levitation-flight trains and vehicles.

## Innovations.

The AB-Levitron uses two large conductivity rings with very high electric currency (fig.1). They create intense magnetic fields. Directions of electric currency are opposed one to the other and rings are repelling one from another. For obtaining enough force over a long distance, the electric currency must be very strong. The current superconductive technology allows us to get very high-density electric currency and enough artificial magnetic field in far space.

The superconductivity ring does not spend an electric energy and can work for a long time period, but it requires an integral cooling system because the current superconductivity materials have the critical temperature about 150-180 C (see Table #1).

However, the present computation methods of heat defense are well developed (for example, by liquid nitrogen) and the induced expenses for cooling are small (fig.2).

The ring located in space does not need any conventional cooling—that defense from Sun and Earth radiations is provided by high-reflectivity screens (fig.3). However, that must have parts open to outer space for radiating of its heat and support the maintaining of low ambient temperature. For variable direction of radiation, the mechanical screen defense system may be complex. However, there are thin layers of liquid crystals that permit the automatic control of their energy reflectivity and transparency and the useful application of such liquid crystals making it easier for appropriate space cooling system. This effect is used by new man-made glasses which grow dark in bright solar light.

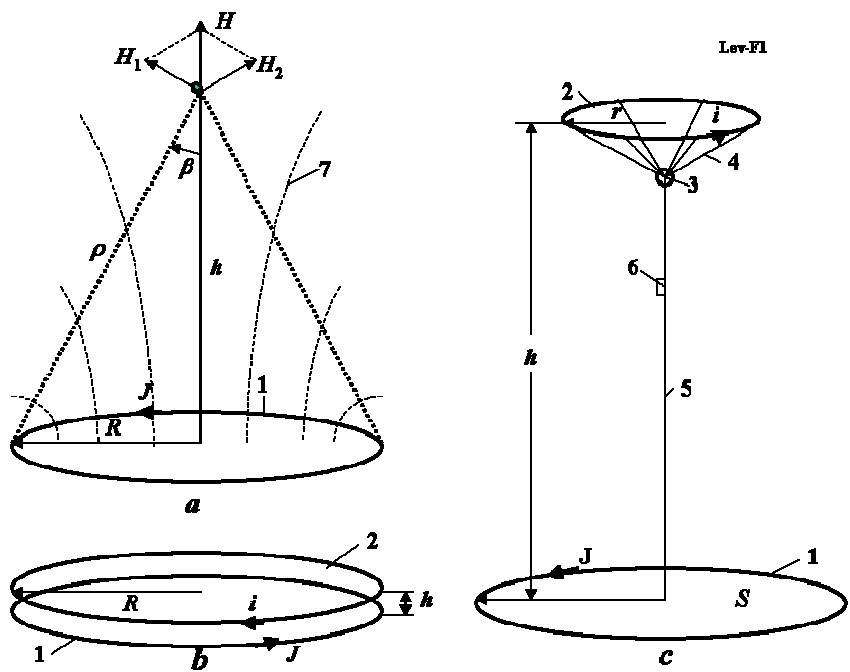

**Fig.1.** Explanation of AB-Levitron. (a) Artificial magnetic field; (b) AB-Levitron from two same closed superconductivity rings; (c) AB-Levitron - motionless satellite, space station or communication mast. Notation: 1- ground superconductivity ring; 2 - levitating ring; 3 - suspended stationary satellite (space station, communication equipment, etc.); 4 - suspension cable; 5 - elevator (climber) and electric cable; 6 - elevator cabin; 7 - magnetic lines of ground ring; $R$ - radius of lover (ground) superconductivity ring; $r$ - radius of top ring; $h$ - altitude of top ring; $H$ - magnetic intensity; $S$ - ring area.

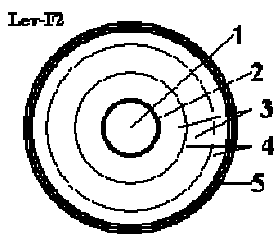

**Fig.2.** Cross-section of superconductivity ring. Notations: 1 - strong tube (internal part used for cooling of ring, external part is used for superconductive layer); 2 - superconductivity layer; 3 - vacuum; 4 – heat impact reduction high-reflectivity screens (roll of thin bright aluminum foil); 5 - protection and heat insulation.

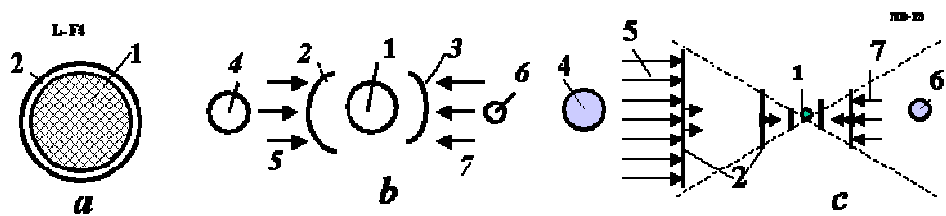

**Fig.3.** Methods of cooling (protection from Sun radiation) the superconductivity levitron ring in outer space. (*a*) Protection the ring by the super-reflectivity mirror [5]. (*b*) Protection by high-reflectivity screen (mirror) from impinging solar and planetary radiations. (*c*) Protection by usual multi-screens. Notations: 1 - superconductive wires (ring); 2 - heat protector (super-reflectivity mirror in Fig.3a and a usual mirror in Fig. 3c); 2, 3 – high-reflectivity mirrors (Fig. 3b); 4 - Sun; 5 -Sun radiation, 6 - Earth (planet); 7 - Earth's radiation.

The most inportant problem of AB-levitron is stability of top ring. The top ring is in equilibrum, but it is out of balance when it is not parallel to the ground ring. Author offers to suspend a load (satellite, space station, equipment, etc)  lower then ring plate. In this case, a center of gravity is lower a summary lift force and system become stable.

For mobile vehicles (fig.7) the AB-Levitron can have a run-wave of magnetic intensity which can move the vehicle (produce electric currency), making it signicantly mobile in the traveling medium.

## Theory of AB-Levitron Estimations, and Computations

1. **Magnetic intensity**. Exactly computation of the magnetic intencity and lift force is complex. We find a simple formula only in two cases: (1) when top ring is small in comparison with ground ring ($r <<$. fig.1c) and located along ground ring axis and (2) the rings are same and closed ($h << R$, fig.1b).
   Results (case 1) are below

$$H = \frac{JS}{2\pi\rho^3} = \frac{JR^2}{2\rho^3}, \quad \rho = (R^2 + h^2)^{1/2}. \quad H = \frac{JR^2}{2(R^2 + h^2)^{3/2}}, \quad B_n = \frac{\mu_0 JR^2}{2(R^2 + h^2)^{3/2}}, \quad (1)$$

where $H$ is magnetic intensity, A/m, along an axis of the ground ring (fig.1a); $J$ is electric currency in the ground ring, A; $S$ is ring area, m$^2$ (fig.1a); $\rho$ is distance from ring element to given point in ring axis, m (fig.1a); $R$ is radius of ground ring, m; $h$ is altitude of top ring, m; $\mu_0 = 4\pi 10^{-7}$ is  magnetic constant, $B_n$ is magnetic intensity which is perpendilar on top ring plate in T.

2. **Lift force**. The lift force is

$$F = p_m \frac{\partial B_n}{\partial h}, \quad p_m = \pi i r^2, \quad F = \pm \frac{3\mu_0 \pi i J r^2 R^2 h}{2(R^2 + h^2)^{5/2}}, \quad (2)$$

where $F$ is lift force, N; $p_m$ is magnetic moment of top ring, A/m$^2$; $i$ is electric currency in top ring, A; $r$ is radius of top ring, m. The sing + or  - depends from direction of electric cirrency in top ring.

3. **Optimal radius of ground ring** for given altitude $h$.  Lift force for given $i, J, r, h$ has maximum

$$A = \frac{R^2 h}{(R^2 + h^2)^{5/2}}, \quad \frac{\partial A}{\partial R} = 0, \quad R_{opt} = \sqrt{\frac{2}{3}} h = 0.8165h, \quad A_{opt} = \frac{0.186}{h^2}, \quad (3)$$

Computation $A$ is presented in fig.4.

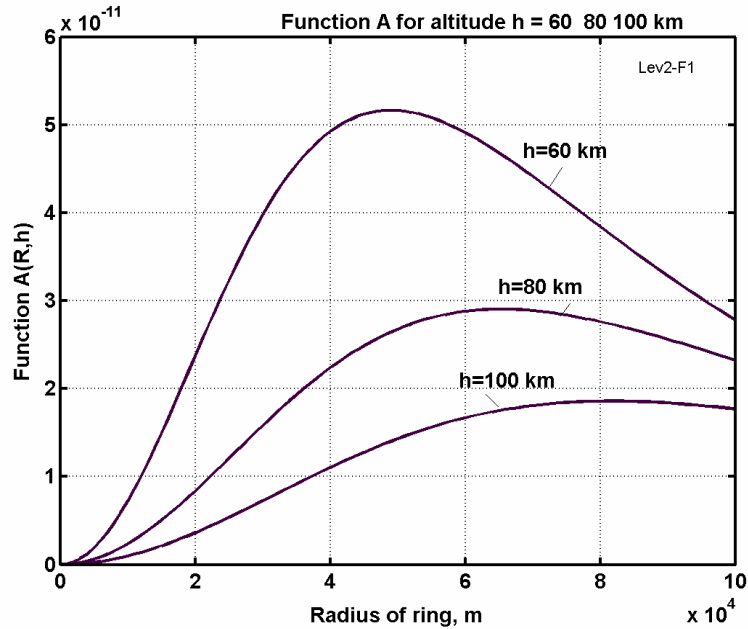

**Fig.4.** Function A versus radius of the ground ring for altitude $h$ = 60, 80, 100 km.

*Note:* For altotude $h$ = 100 km, the optimal radius is $R_{opt}$ = 81.65 km. However, the decreasing this radius from 81.65 to 65 km decreases the lift force only in 5% (fig.4).

The magnetic intensity and force corresponding the $R_{opt}$ are

$$B_{n,R\,opt} \approx \frac{\mu_0 J}{11.86h}, \quad F_{R\,opt} \approx \frac{\pi\mu_0 iJ}{10\bar{h}^2}, \quad \text{where} \quad \bar{h} = \frac{h}{r}, \tag{4}$$

*Example:* If $i = 10^7$ A, $J = 10^9$ A, $\bar{h} = 10$, then $F = 4\times10^6$ N = 400 tons. If the $h$ = 100 km that means the $R = 65 \div 81$ km, $r$ = 10 km.

Computation of lift force for $R_{opt}$ and relative altitude $\bar{h} = 10$ is presented in fig.5

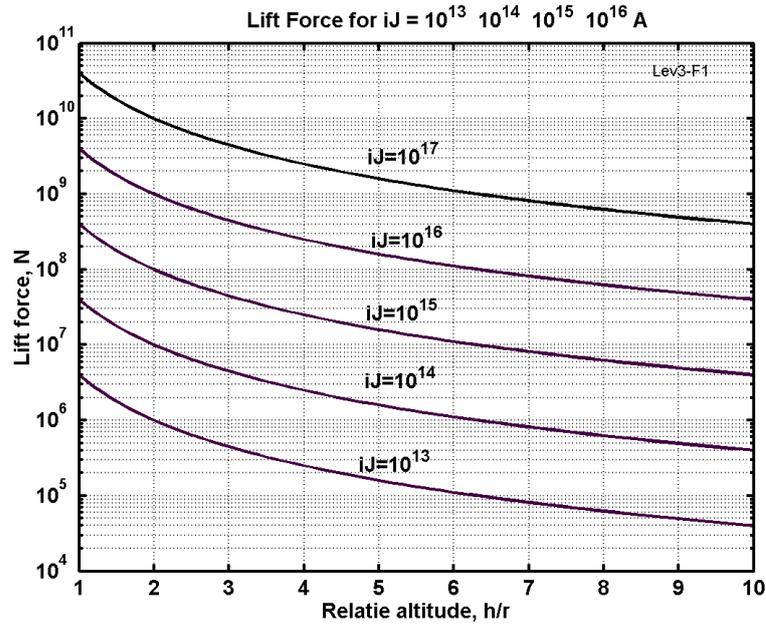

**Fig. 5.** Lift force of AB-Levitron space station versus the relative altitude for a product of the electric currencies the superconductivity ground and station rings and for the optimal ground ring. $\bar{h} = h/r$, $h$ is station altitude, $r$ is radius of top ring.

**4. The lift force in case (2) (fig.1b).** In this case the lift force is

$$F = \pm\mu_0 i J \frac{R}{h}, \tag{5}$$

**6. The lift force in case of linear AB-highway** (fig. 7). This lift force can be estimated by equation ($h \ll L$)

$$F = \mu_0 i J \frac{L}{2\pi h}, \quad \text{for } L = 1 \text{ m}, \quad F_1 = \mu_0 i J \frac{1}{2\pi h} = \frac{2\cdot 10^{-7} iJ}{h}, \tag{6}$$

where $L$ is length of AB-train (vehicle), m; $F_1$ is lift force the 1 m length of train (vehicle). The computation is presented in fig. 6.

**7. Lift force in general case.** This lift force of the top ring can be computed by equation

$$\bar{F} = (\bar{p}_m \text{grad})\bar{B} \quad \text{or for axis } x \quad F_x = p_{mx}\frac{\partial B_x}{\partial x} + p_{my}\frac{\partial B_x}{\partial y} + p_{mz}\frac{\partial B_x}{\partial z}. \tag{7}$$

where $p_m = iS_t$ is magnetic moment, N·m; $S_t$ is area of top ring, m².

**8. Some other parameters.** The moment of force in the top ring is

$$M = [\bar{p}_m \cdot \bar{B}]. \tag{8}$$

When the currency in ground ring is variable, the voltage and electric currency in top ring are

$$\mathrm{E} = -\frac{d\Phi}{dt}, \quad i = -\frac{E}{r_t}, \quad \text{where} \quad \Phi = S_t B_n, \tag{9}$$

where E is voltage induced in top ring, V; $\Phi$ is magnetic flow throw the top ring, Wb; $r_t$ is electric resistance of the top ring, Ω.

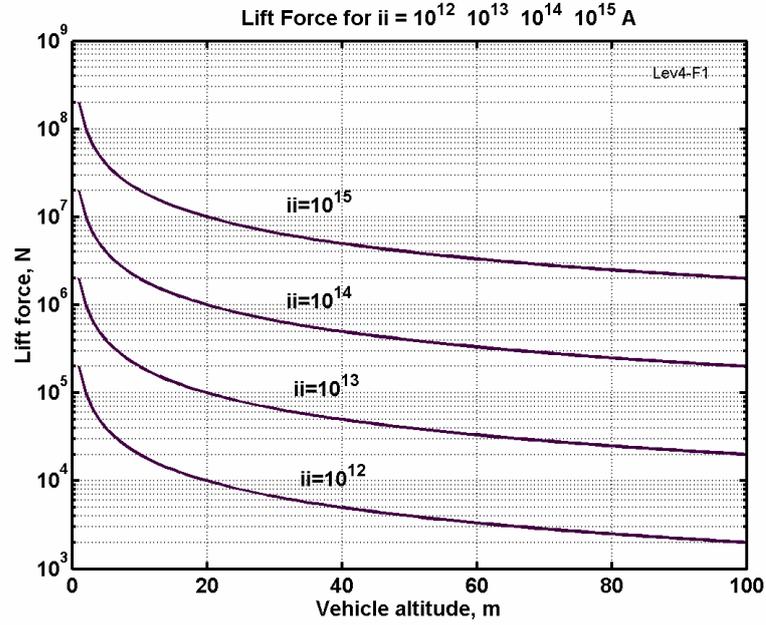

**Fig. 6.** Lift force [N/m] of the 1 meter AB-Levitron multipath highway versus the train (vehicle) altitude for a product (*ii*) of the electric currencies of the superconductivity ground and vehicle rings.

The minimal radius $R_{T,min}$ [m] of the ring tube and a maximal magnetic pressure $P_{T,max}$ [N/m²] are

$$R_{T,min} = \frac{\mu_0 i}{2\pi B}, \quad P_{T,max} = \frac{B^2}{2\mu_0}, \quad P_T = \frac{\mu_0 i^2}{8\pi^2 R_T^2}, \tag{10}$$

where *B* is maximum safety magnetic intensity for given superconductivity material, T (see Table #1).

*Example:* for $i = 10^7$ A, $B = 100$ T we have $R_{T,min} = 5$ mm, $P_{T,max} = 4 \times 10^9$ N/m² = $4 \times 10^4$ atm.

The pressure is high. Steel 40X has a limit $4 \times 10^9$ N/m², corundum has a limit $21 \times 10^9$ N/m². However, we can adopt a larger tube radius $R_T$ and, as a result, then decrease the magnetic pressure. The internal cooling gas also has pressure which is opposed the magnetic pressure.

**9. Energy of superconductivity ring**, If the magnetic intensity into ring is constant, we can estimate the energy needed for starting of ring:

$$H = \frac{I}{2R}, \quad \Phi = \mu_0 \frac{I}{2R} S = IL, \quad L = \mu_0 \frac{S}{2R} = \mu_0 \frac{\pi R}{2}, \quad E = \frac{LI^2}{2} = 0.25\pi\mu_0 RI^2, \tag{11}$$

where $\Phi$ is magnetic flux, Wb: *L* is ring inductance, Henry; *S* is ring area, m²; final equation in (11) *E* is energy, J; *I* is electric currency, A.

*Example:* For ground ring having $R = 10$ km and $I = 10^8$ A the $E = 10^{14}$ J = $2.5 \times 1000$ tons of fuel (gas having specific energy $40 \times 10^6$ J/kg). For top ring having $R = 100$ m and $I = 10^6$ A the $E = 10^8$ J = 2.5 kg of fuel (gas).

As the reader will undoubtedly readily note, the superconductivity ground ring is an excellent storage of electric energy.

**10. Ring internal pressure** is

$$f_r = \frac{\mu_0 H^2}{2}, \quad H = \frac{i}{2r}, \quad f_r = \frac{\mu_0 i^2}{8r^2}, \tag{12}$$

In our macro-projects (for large *r*) this pressure is small.

**11. Mass of suspension cables** $m_s$ when $m_s \ll M_S$, $m_s \ll M_r$

$$m_s = M_S \frac{2gr\gamma}{\sigma \sin 2\alpha}, \qquad (13)$$

where $M_S$ is space station mass, kg; $M_r$ is top ring mass, kg; $g = 9.81$ m/s² is gravity; $\gamma$ is specific mass of suspended cable, kg/m³; $\sigma$ is safety tensile stress of suspended cable, N/m²; $\alpha$ is angle between plate of top ring and the suspended cable.

**12. Minimal rotation speed of top ring** for keeping of space station (when $M_r \gg m_s$)

$$V = \sqrt{\frac{2grM_S}{M_r \sin 2\alpha}}, \quad t = \frac{2\pi r}{V} \qquad (14)$$

where $V$ is rotation speed of top ring, m/s; $t$ is time of one revolution, sec.

**13. Superconductivity materials.**

There are hundreds of new superconductivity materials (type2) having critical temperature 70 ÷ 120 K and more.

*Some of the superconductable materials* are presented in Table 1 (2001). The widely used YBa$_2$Cu$_3$O$_7$ has mass density 7 g/cm³.

**Table 1**. Transition temperature $T_c$ and upper critical field $B=H_{c2}(0)$ of some examined superconductors [7], p. 752.

| Crystal | $T_c$ (K) | $H_{c2}$ (T) |
|---|---|---|
| La $_{2-x}$Sr$_x$CuO$_4$ | 38 | ≥80 |
| YBa$_2$Cu$_3$O$_7$ | 92 | ≥150 |
| Bi$_2$Sr$_2$Ca$_2$Cu$_3$O$_{10}$ | 110 | ≥250 |
| TlBa$_2$Ca$_2$Cu$_3$O$_9$ | 110 | ≥100 |
| Tl$_2$Ba$_2$Ca$_2$Cu$_3$O$_{10}$ | 125 | ≥150 |
| HgBa$_2$Ca$_2$Cu$_3$O$_8$ | 133 | ≥150 |

The last decisions are: Critical temperature is 176 K, up 183 K. Nanotube has critical temperature 12 - 15 K,

Some organic matters have a temperature of up to 15 K. Polypropylene, for example, is normally an insulator. In 1985, however, researchers at the Russian Academy of Sciences discovered that as an oxidized thin-film, polypropylene have a conductivity $10^5$ to $10^6$ that is higher than the best refined metals.

Boiling temperature of liquid nitrogen is 77.3 K, air 81 K, oxygen 90.2 K, hydrogen 20.4 K, helium 4.2 K [8].

Unfortutately, most superconductive material is not strong and needs a strong covering.

**14. Computation of the cooling system**. The following equations allow direct computation of the proposed macro-project cooling systems.
1) Equation of heat balance of a body in vacuum

$$\zeta q s_1 = C_s \varepsilon_a \left(\frac{T}{100}\right)^4 s_2, \qquad (15)$$

where $\zeta = 1 - \xi$ is absorption coefficient of outer radiation, $\xi$ is reflection coefficient; $q$ is heat flow, W/m² (from Sun at Earth's orbit $q = 1400$ W/m², from Earth $q \approx 440$ W/m²); $s_1$ is area under outer radiation, m²; $C_s = 5.67$ W/m²K is heat coefficient; $\varepsilon_a \approx 0.02 \div 0.98$ is blackness coefficient; $T$ is body temperature, K; $s_2$ is area of body or screen, m².

2) Radiation heat flow $q$ [W/m²] between two parallel screens

$$q = C_a \left[\left(\frac{T_1}{100}\right)^4 - \left(\frac{T_2}{100}\right)^4\right], \quad C_a = \varepsilon_a C_s, \quad \varepsilon_a = \frac{1}{1/\varepsilon_1 + 1/\varepsilon_2 - 1}, \qquad (16)$$

where the lower index $_{1,2}$ shows (at $T$ and $\varepsilon$) the number of screens; $C_a$ is coerced coefficient of heat transfer between two screens. For bright aluminum foil $\varepsilon = 0.04 \div 0.06$. For foil covered by thin bright layer of silver $\varepsilon = 0.02 \div 0.03$.

When we use a vacuum and row ($n$) of the thin screens, the heat flow is

$$q_n = \frac{1}{n+1} \frac{C_a'}{C_a} q, \qquad (17)$$

where $q_n$ is heat flow to protected wire, W/m$^2$; $C_a'$ is coerced coefficient of heat transfer between wire and the nearest screen, $C_a$ is coerced coefficient of heat transfer between two near by screens; $n$ is number of screen (revolutins of vacuumed thin foil around central superconductive wire).

*Example*: for $C_a' = C_a$, $n = 100$, $\varepsilon = 0.05$, $T_1 = 298$ K (15 C, everage Earth temperature), $T_2 = 77.3$ K (liquid nitrogen) we have the $q_n = 0.114$ W/m$^2$.

Expence of cooling liquid and power for converting back the vapor into cooling liquid are

$$m_a = q_n / \alpha, \quad P = q_n S / \eta, \qquad (18)$$

where $m_a$ is vapor mass of cooling liquid, kg/m$^2$.sec; $P$ is power, W/m$^2$; $S$ is an outer area of the heat protection, m$^2$; $\eta$ is coefficient of efficiency the cooling instellation which convert back the cooling vapor to the cooling liquid; $\alpha$ is heat varoparation, J/kg (see Table 2).

3) When we use the conventional heat protection, the heat flow is computed by equations

$$q = k(T_1 - T_2), \quad k = \frac{\lambda}{\delta}, \qquad (19)$$

where $k$ is heat transmission coefficient, W/m$^2$K; $\lambda$ - heat conductivity coefficient, W/m·K. For air $\lambda = 0.0244$, for glass-wool $\lambda = 0.037$; $\delta$ - thickness of heat protection, m.

The vacuum screenning is strong efficiency and light (mass) then the conventional cooling protection.

**Table 2**. Boiling temperature and heat of evaporation of some relevant liquids [8]. p.68 .

| Liquid | Boilng temperature, K | Heat varoparation, $\alpha$ kJ/kg |
|---|---|---|
| Hydrogen | 20.4 | 472 |
| Nitrogen | 77.3 | 197.5 |
| Air | 81 | 217 |
| Oxygen | 90.2 | 213.7 |
| Carbonic acid | 194.7 | 375 |

These data are sufficient for a quick computation of the cooling systems characteristics.

Using the correct design of multi-screens, high-reflectivity solar and planetary energy screen, and assuming a hard outer space vacuum between screens, we get a very small heat flow and a very small expenditure for refrigerant (some gram/m$^2$ per day in Earth). In outer space the protected body can have low temperature without special liquid cooling system (Fig.3).

For example, the space body (Fig. 3a) with innovative prism reflector [5] Ch. 3A ($\rho = 10^{-6}$, $\varepsilon_a = 0.9$) will have temperature 13 K in outer space. The protection Fig.3b gives more low temperature. The usual multi-screen protection of Fig. 3c gives the temperature: the first screen - 160 K, the second - 75 K, the third - 35 K, the fourth - 16 K.

**15. Cable material**. Let us consider the following experimental and industrial fibers, whiskers, and nanotubes:
1. Experimental nanotubes CNT (carbon nanotubes) have a tensile strength of 200 Giga-Pascals (20,000 kg/mm$^2$). Theoretical limit of nanotubes is 30,000 kg/mm$^2$.

2. Young's modulus exceeds a Tera Pascal, specific density $\gamma = 1800$ kg/m$^3$ (1.8 g/cc) (year 2000).
   For safety factor $n = 2.4$, $\sigma = 8300$ kg/mm$^2$ = $8.3 \times 10^{10}$ N/m$^2$, $\gamma = 1800$ kg/m$^3$, $(\sigma/\gamma) = 46 \times 10^6$. The SWNTs nanotubes have a density of 0.8 g/cm$^3$, and MWNTs have a density of 1.8 g/cm$^3$ (average 1.34 g/cm$^3$). Unfortunately, even in 2007 AD, nanotubes are very expensive to manufacture.
3. For whiskers $C_D$ $\sigma = 8000$ kg/mm$^2$, $\gamma = 3500$ kg/m$^3$ (1989) [5, p. 33]. Cost about $400/kg (2001).
4. For industrial fibers $\sigma = 500 - 600$ kg/mm$^2$, $\gamma = 1800$ kg/m$^3$, $\sigma/\gamma = 2{,}78 \times 10^6$. Cost about 2 - 5 $/kg (2003).

Relevant statistics for some other experimental whiskers and industrial fibers are given in Table 3 below.

Table 3. Tensile strength and density of whiskers and fibers

| Material Whiskers | Tensile strength kg/mm$^2$ | Density g/cm$^3$ | Fibers | Tensile strength kg/mm$^2$ | Density g/cm$^3$ |
|---|---|---|---|---|---|
| AlB$_{12}$ | 2650 | 2.6 | QC-8805 | 620 | 1.95 |
| B | 2500 | 2.3 | TM9 | 600 | 1.79 |
| B$_4$C | 2800 | 2.5 | Thorael | 565 | 1.81 |
| TiB$_2$ | 3370 | 4.5 | Alien 1 | 580 | 1.56 |
| SiC | 2100-4140 | 3.22 | Alien 2 | 300 | 0.97 |
| Al oxide | 2800-4200 | 3.96 | Kevlar | 362 | 1.44 |

See Reference [5] p. 33.

16. **Safety of space station.** For safety of space station and elevator cabin the special parachutes are utilized (see [9]). Author also has ideas for the safety of the ground superconductivity ring.

## Projects
### Macro-Project #1. Stationary space station at altitude 100 km.

Let us to estimate the stationary space station is located at altitude $h = 100$ km. Take the initial data: Electric currency in the top superconductivity ring is $i = 10^6$ A; radius of the top ring is $r = 10$ km; electric currency in the superconductivity ground ring is $J = 10^8$ A; density of electric currency is $j = 10^6$ A/mm$^2$; specific mass of wire is $\gamma = 7000$ kg/m$^3$; specific mass of suspending cable and lift (elevator) cable is $\gamma = 1800$ kg/m$^3$; safety tensile stress suspending and lift cable is $\sigma = 1.5 \times 10^9$ N/m$^2$ = 150 kg/mm$^2$; $\alpha = 45^\circ$, safety superconductivity magnetic intensity is $B = 100$ T. Mass of lift (elevator) cabin is 1000 kg.

Then the optimal radius of the ground ring is $R = 81.6$ km (Eq, (3), we can take $R = 65$ km); the mass of space station is $M_S = F = 40$ tons (Eq.(2)). The top ring wire mass is 440 kg or together with control screen film is $M_r = 600$ kg. Mass of two-cable elevator is 3600 kg; mass of suspending cable is less 9600 kg, mass of parachute is 2200 kg. As the result the useful mass of space station is $M_u = 40 - (0.6+1+3.6+9.6+2.2) = 23$ tons.

Minimal wire radius of top ring is $R_T = 2$ mm (Eq. (10)). If we take it $R_T = 4$ mm the magnetic pressure will me $P_T = 100$ kg/mm$^2$ (Eq. (10)). Minimal wire radius of the ground ring is $R_T = 0.2$ m (Eq. (10)). If we take it $R_T = 0.4$ m the magnetic pressure will me $P_T = 100$ kg/mm$^2$ (Eq. (10)). Minimal rotation speed (take into consideration the suspending cable) is $V = 645$ m/s, time of one revolution is $t = 50$ sec. Electric energy in the top ring is small, but in the ground ring is very high $E = 10^{14}$ J (Eq. (11)). That is energy of 2500 tons of liquid fuel (such as natural gas, methane).

The requisite power of the cooling system for ground ring is about $P = 30$ kW (Eq. (18)).

As the reader observes, all parameters are accessible using existing and available technology. They are not optimal.

## Macro-Project #2. 500 m-high Tele-communication Mast.

Let us estimate the tele-communication mast of height $h$ = 500 m without superconductivity in the top ring. Take the initial data: Electric currency in the top ring is $i$ = 100 A; radius of the top ring is $r$ = 200 m; electric currency in the superconductivity ground ring is $J$ = 2.5×10$^8$ A; density of ground ring electric currency is $j$ = 10$^6$ A/mm$^2$, the top ring has $j$ = 5 A/mm$^2$; specific mass of superconductivity wire is $\gamma$ = 7000 kg/m$^3$; specific mass of aluminum wire is $\gamma$ = 2800 kg/m$^3$; specific mass of suspending cable and lift cable is $\gamma$ = 1800 kg/m$^3$; safety tensile stress suspending and list cable is $\sigma$ = 10$^9$ N/m$^2$ = 100 kg /mm$^2$; $\alpha$ = 45°, safety superconductivity magnetic intensity is $B$ = 100 T. The vertical wire transfer of electric energy has voltage 2000 V and electric density 8.8 A/mm$^2$. Then the optimal radius of the ground ring is $R$ = 400 m (Eq. (3)); the mass of antenna is $M_S = F$ = 160 kg (Eq.(2)). The top ring wire mass is $M_r$ = 70 kg. Mass of vertical two-cable transfer of electric energy is 3 kg; mass of suspending cable is less 1 kg. As the result the useful mass of top apparatuses is $M_u$ = 160 - (70+3+1) = 86 kg.

Minimal wire radius of ground ring is $R_T$ = 0.5 m (Eq. (10)). If we take it $R_T$ = 1.5 m the magnetic pressure will me $P_T$ = 44 kg/mm$^2$ (Eq. (10)). Minimal rotation speed of top ring is $V$ = 96 m/s, time of one revolution is $t$ = 12.6 sec. Electric energy in the top ring is small, but in the ground ring is high $E$ = 2.5×10$^{13}$ J (Eq. (11)). That is energy of 620 tons of liquid fuel (natural gas). Requested energy for permanent supporting the electric currency in NON-SUPERCONDUCTIVITY top ring is 17.6 kW. If we make it superconductive, the lift force increases by thousands times. For example, if $i$ = 10$^6$ A the lift force increases in 10$^6$/100 = 10$^4$ times and became 1.6×10$^3$ tons. That is suspending mobile building (hotel). There is no expense of electric energy for superconductivity ring. The power for cooling (liquid nitrogen) is small. It is not used an expensive city area.

As it is shown in the author work [4] we can build the fight city where men can fly as individuals and also in the cars or similar vehicles.

All parameters are accessible for existing industry. They are not optimal. Our aim - it shows that AB-Levitron may be designed by the current technology (see also [11]).

## Macro-Project #3. Levitron AB-miltipath highway.

The AB-levitron may be used for design the multi-path levitation highway (fig.7). That is the closed-loop superconductive lenthy linear strung near the highway which creates the vertical magnetic field. The lift force produced by this AB-highway in one meter of length is [Eq. (6)]

$$F_1 = \frac{2 \cdot 10^{-7} i_1 i_2}{h}, \quad (17)$$

where $F_1$ is lift force, N/m; $i$ is electric currency in graund cable and fly train (vehicle) respectively; $h$ is altitude of the veficle (train) over graund cable, m.

*Estimations*. Let us take the electric currency $i_1$ = 10$^8$ A in ground line. Then:
1) If the the train does not have the superconductivity wire, the electric currency in top ring is only $i_2$ = 100 A and distance between rings is $h$ = 0.5 m, the lift force of 1 m train length will be $F_1$ = 4000 N/m (Eq. (6 or 17)). In this case the top ring may be changed by permanent magnets.
2) If the vehicle has the supercondutivity with currency $i_2$ = 10$^6$ A then the 1 m length of vehicle will has:
    a) at altitude $h$ = 10 m the $F_1$ = 200 ton/m ;
    b) at altitude $h$ = 100 m the $F_1$ = 20 ton/m ;
    c) at altitude $h$ = 1000 m the $F_1$ = 2 ton/m ;

If vertical distance between paths is 10 meter the 1 km vertical corridor will have 100 ways in one direction from a lower low speed vehicles to a top high speed vehicles (supersonic aircraft). They can receive energy from running magnet wave of graund cable.

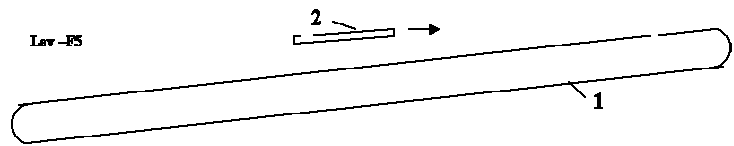

**Fig.7**. High speed AB-Levitron way for levitating aircraft or trains. Notations: 1 - ground superconductivity closed-loop cable; 2 - car, track, air vehicle or train.

The offered AB-Llevitron vehicle has following advantages in comparison the current train used magnet pillow (maglev):
1) One does not need in complex expencive magnet system;
2) That does not need a precise concrete roadbed;
3) There area a lot of ways and train can change a path and it does not depend from condition of other train and vehicles,

The initial data in all our macro-projects are not optimal.

## Discussion

The offered AB-Levitrons may be made with only existing technology. We have a superconductivity material (see Table 1), the strong artificial fibers and whiskers (Table 3), the light protection and cooling system (Table 2) for the Earth's surface, and the radiation screens for outer space. The Earth has weak magnetic field, the Sun and many planets and their satellites (as Phobos orbiting Mars) has also small magnetic field. There is no barrier problem to creating the artificial magnetic field on Earth, asteroids and planetary satellites (for example, to create local artificial magnetic field on the Moon, see [10]). We have a very good perspective in improving our devices because—especially during the last 30 years—the critical temperature of the superconductive material increases from 4 K to 186 K and does not appear, at this time, to be any theoretical limit for further increase. Moreover, Russian scientists received the thin layers which have electric resistance at room temperature in many times less than the conventional conductors. We have nanotubes which will create the jump in AB-Levitrons, when their production will be cheaper. The current superconductive solenoids have the magnetic field $B \approx 20$ T.

AB-levitrons can instigate a revolution in space exploration and exploitation, tele-communication and air, ground, and space vehicle transportation. They allow individuals to fly as birds, almost flight with subsonic and supersonic speed [4]. The AB-Levitrons solve the environment problem because they do not emit or evolve any polluting gases. They are useful in any solution for the national and international oil-dependence problem because they use electricity and spend the energy for flight and other vehicles (cars) many times less than conventional internal combustion engine (no graund friction). In difference of a ground car, the levitation car flights are straight line to objective in a city region.

The AB-Levitrons create a notable revolution in tele-communication by the low-altitude stationary suspended satellites, in energy industry, and especially in a local aviation. They are very useful in night-lighting of Earth-biosphere regions by additional light and heat Sun radiation because, in difference from conventional mobile space mirrors, they can be suspended over given place (city) and service this place efficiently.

It is interesting, the toroidal AB engine is very comfortable for flying discs (human-made UFO!) and have same property with UFOs. That can levitate and move in any direction with high acceleration without turning of vehicle, that does not excrete any gas, jet, and that does not produce a noise [4].

## Conclusion

We must research and develop these ideas as soon as possible. They may accelerate the technical progress and improve our life-styles. There are no known scientific obstacles in the development and design of the AB-Levitrons, levitation vehicles, high-speed aircraft, spaceship launches, low-aititude stationary tele-communication satellites, cheap space trip to Moon and Mars and other interesting destination-places in outer space.

## Acknowledgement

The author wishes to acknowledge R.B. Cathcart for correcting the author's English and some useful advice.